\documentstyle[a4,12pt]{article}

\def\be{\begin{equation}}
\def\ee{\end{equation}}
\def\bea{\begin{eqnarray}}
\def\eea{\end{eqnarray}}

\def\re#1{(\ref{#1})}
\def\year{{\rm{year}}}
\def\bpyear{{\rm{[b.p./year]}}}
\def\yearbp{{\rm{[year/b.p.]}}}
\def\century{{\rm{[century]}}}
\def\pcyear{{\rm{[\%/year]}}}
\def\yearpc{{\rm{[year/\%]}}}
\def\timedim{{\rm{[[t]]}}}
\def\bpy{[b.p./year]}
\def\ybp{[year/b.p.]}
\def\pcy{[\%/year]}
\def\ypc{[century]}
\def\td{$[[t]]$}
\def\half{$^{\frac{1}{2}}$}

\def\LL{$L$}
\def\ss{$s$}
\def\qu{$q_1$}
\def\qd{$q_2$}
\def\dqu{$\Delta q_1$}
\def\dqd{$\Delta q_2$}
\def\ds{$\Delta s$}
\def\C{$\chi^2$}

\def\matur#1{$[m]=[#1]$}
\def\laqu{$\lambda_{q_1}$}
\def\laqd{$\lambda_{q_2}$}
\def\nuqu{$\nu_{q_1}$}
\def\nuqd{$\nu_{q_2}$}
\def\dlaqu{$\Delta\lambda_{q_1}$}
\def\dlaqd{$\Delta\lambda_{q_2}$}
\def\dnuqu{$\Delta\nu_{q_1}$}
\def\dnuqd{$\Delta\nu_{q_2}$}

\begin{document}
\begin{titlepage}
\rightline{cond-mat/9901096}
\vskip 1.5 true cm
\large{\centerline{ Phenomenology of the Term Structure }}
\vskip 0.5 true cm
\large{\centerline{of Interest Rates with Pad\'e Approximants      
\footnote{The authors would like to thank Professor Virasoro
(ASICTP) and Professor Amati (SISSA) for hospitality in Trieste
while part of this paper was being written and Professor
Rittenberg for his remarks and encouragements.
}}}
\vskip 0.6 true cm
\large{\centerline{ Jean Nuyts$^a$\ \ and\ \   Isabelle Platten$^b$}}
\vskip 0.5 true cm
\small{\centerline{Jean.Nuyts@umh.ac.be, Isabelle.Platten@fundp.ac.be
\footnote{Corresponding author.
}}}
\vskip 0.2 true cm
\centerline{\it $^a$  Physique Th\'eorique et Math\'ematique, 
Universit\'e de Mons-Hainaut}
\centerline{\it 20 Place du Parc, 7000 Mons, Belgium}
\centerline{\it $^b$ CeReFIM, 
Facult\'es Universitaires Notre-Dame de la Paix}
\centerline{\it 8 Rampart de la Vi\`erge, 5000 Namur, Belgium}
\vskip 1.3 true cm
\noindent{\bf Abstract}
\small{
\begin{quote}
The classical approach in finance attempts to model the term
structure of interest rates using specified stochastic processes
and the no arbitrage argument. 
Up to now, no universally accepted theory has been
obtained for the description of
experimental data. 
We have chosen
a more phenomenological approach. It is based on results
obtained some twenty years ago by physicists, results which show
that Pad\'e Approximants are very suitable for approximating large
classes of functions in a very precise and coherent way. In this
paper, we have chosen to compare Pad\'e Approximants with very low
indices with the experimental densities of interest rates variations.
We have shown that the data
published by the Federal Reserve System in the United States
are very well reproduced with two
parameters only. These parameters are rather simple
functions of the lag and of the maturity and are directly
related to the moments of the distributions. 
\end{quote}
      }
\vfill
\end{titlepage}

\section{Introduction}

The classical approach in finance attemps to model the term
structure of interest rates using specified stochastic processes
and the no arbitrage argument. 
Numerous theoretical models (see for example  \cite{V}, \cite{CIR},
\cite{HJM}, \cite{HW}) have been proposed.
Although they provide analytical formulas for
the pricing of interest rate derivatives, the implied
deformations of the term structure 
have a Brownian motion component and are often rejected by
empirical data \cite{CKLS}.

The inadequacies of the Gaussian model for the description of
financial time series has been reported for a long time
\cite{M1}, \cite{M2} but the availability of enormous sets of
financial data, with a small time scale, has renewed the
interest in the subject (see \cite{MDOPSM}, 
\cite{MS94}, \cite{GBPTD}, \cite{MS97}).
In particular, the
fat-tail property of the empirical distributions of price
changes has been widely documented and is a crucial feature for
monitoring the extreme risks \cite{EKM}, \cite{J}. 
Most of the
recent studies concern stock indexes or exchange rates
\cite{MS97}, \cite{GBPTD}, \cite{MDOPSM}, with high frequency data. 
Studies on
interest rates are rather rare and often limited to a few
maturities (three and six months cash rates in \cite{MDP}, Bund
futures in \cite{BP}).
The paper \cite{BSCEP}
is an exception as the US forward rate curve with maturities up to four
years is modeled.

There is a long history of attemps to model price fluctuations
with Levy distributions \cite{M1}. The family of Levy laws
exhibits qualitatively heavy tails and enjoys the suitable
properties of stable laws \cite{BP}, \cite{F}.
However, they are defined in an unduly complicated manner (using
the characteristic function) and do not even have a finite
moment of order two (variance). Truncated Levy distributions
have been proposed to circumvent this drawback \cite{MS94},
\cite{Mata}.  Other alternative distributions have also been
proposed (see \cite{K}, \cite{EKP}, \cite{MDDOPW}).

Our aim, in this paper, is to propose an alternative description
of the empirical distributions of the variations of interest
rates in a simple form, with a small number of parameters but
enjoying the property of providing a remarkable fit of the full
distribution and hence, at the same time, of both the
central part and the tails. 

Twenty five years ago, physicists have started studying
sets of reduced rational forms which are particularly 
well suited to approximate, in a very uniform way, 
functions which belong to rather large classes. These
forms are called Pad\'e Approximants. 
Many major progresses in
the understanding of these approximants have been made at that time.
Since there is a very extensive literature concerning the
validity and the usefulness of these approximants, which are
intrinsically related to the theory of rational polynomials, to
the theory of
continuous fractions and to the theory of moments, we refer the
reader directly to some relevant books 
\cite{G}, \cite{A}, \cite{B} where more references can be found. 
We restrict ourselves, here, to
a short presentation of the ideas leaving out the fine points.
The basic idea of the Pad\'e Approximants is to approximate a
large class of functions of, say, the variable $v$
by the ratio of two polynomials $T^M(v)$ and $B^N(v)$ of
respective degrees $M$ and $N$. 

In this paper, we propose to apply the 
Pad\'e Approximants to the study of
daily interest rate changes and show their relevance on data
obtained from the American bond market from 1977 to 1997 \cite{FRS}. 

In Section~2, we discuss the discreteness of the data and their
approximation by continuous distributions. In Section~3, some
general properties of the Pad\'e Approximants are outlined.
Section 4 contains a justification of the criteria chosen in
order to assess the quality of the fit. The data are analyzed and
our results are presented in Section~5. Section~6 contains the
conclusions and some final remarks.

\section{Interest Rates} 

Consider the term structures of interest rates given by
samples of $N_{tot}$ daily interest rates for
constant maturities $[m]$. 
Let us call
\be
I^{[m]}(t),\ \ t=1,\ldots,N_{tot}
\,,
\label{interestrates}
\ee
these spot interest rates where the upper index $[m]$ 
(in unit $\year$) specifies the maturity
and $t$ indexes, chronologically, the opening days.

A short digression on our notation for the ``units'' of interest rates 
is useful at this point.
Interest rates are canonically, 
like in the data base we used,
given in ``percent/year'' (which we will
henceforth denote $[\%/\year]$ and
which is usually and abusively simply called ``percent'') with
two meaningful digits after the decimal point. Hence, the
natural unit we will use 
for interest rate variations is the 
basis point $10^{-4}/\year$ which we denote $\bpyear$. 
The inverse unit is $\yearbp$  will also be used.
The daily changes in interest rates 
are obviously also recorded as integers in units $\bpyear$
and vary from roughly $-100$ to $+100$. 
for a maturity of one year
(this interval may be larger for higher lags).
The unit $\bpyear$  is the natural unit to perform discrete
summations and to expose the problem most simply.
However, at the end, the parameters entering the problem  and
determined from the data are
recorded in the $\pcyear$ (and its inverse unit) better
adapted to their actual values.

See Appendix A for a detailed discussion of the dimensions of
the variables and of the parameters we will use and hence of their
behaviour when units are changed.
 
\subsection{Variations of the Interest Rates} 

From the interest rates $I^{[m]}(t)$ (integers in units $\bpyear$), variations 
of the interest rates $\delta I^{[m]}_L(t)$
at a lag of
$L$ days can be defined for each maturity $[m]$ as
\be
\delta I^{[m]}_L(t)= I^{[m]}_L(t+L) - I^{[m]}_L(t),
\ \ t=1,\ldots,(N_{tot}-L) 
\,.
\ee
These (overlapping) variations of interest rates 
are expressed again as integers in units $\bpyear$. 
Our main goal is to study and to compare the distributions of 
$\delta I^{[m]}_L(t)$ for all available maturities and for all
time lags which we have chosen in the range from $L=1$ to $L=30$
days.   

\subsection{Distribution of the Variations of the Interest Rates} 

For every variation equal to a given value $\hat v$ 
in units $\bpyear$ (where $\hat v$
is an integer), we can
count the number of times this variation occurs 
in the experimental sample. We denote these distributions 
of variations by $N^{[m]}_L(\hat v)$
\bea
N^{[m]}_L(\hat v) &\equiv&\ {\rm{number\ of\ times}}
                       \ \delta I^{[m]}_L(t)=\hat v
                       \nonumber \\            
   &=&\ {\rm{card}} \ \left\{ t, \ \delta I^{[m]}_L(t)=\hat v \right\}
\,.
\label{bareN}
\eea
By construction, we have $N^{[m]}_L(\hat v)$ equal to 
zero outside the range $[\hat v_{min},\hat v_{max}]$
\bea
\hat v_{min}&=&\min_t\left\{  \delta I^{[m]}_L(t)  \right\}
      \nonumber \\
\hat v_{max}&=&\max_t\left\{  \delta I^{[m]}_L(t)  \right\}
\label{vminmax}
\eea
and
\be
\sum_{\hat v=\hat v_{min}}^{\hat v_{max}} N^{[m]}_L(\hat v)=N_{tot}-L
\,.
\label{normbareN}
\ee 
Obviously, $\hat v_{min}$ and $\hat v_{max}$ depend on the maturity $[m]$
and on the lag $L$ but, to simplify the formulas, we have
suppressed these extra labels. 

At this point, it is worth making two comments. 
\begin{enumerate}

\item The opening days are not always consecutive, due to the
presence of Sundays and other non working days. We have
considered that the definition of the lag ignores these gaps. 
We work with a business time scale rather than a physical one.
A more careful study could
take into account this fact and limit itself to consider as
belonging to lag one only those days which are non separated by
one or more missing days. In order to keep our useful data as
large as possible we have chosen to ignore this difficulty. A
more refined discussion could obviously take this into account
at the price of a lowering in statistics.
With high frequency data, the
definition of a time scale that models the market activity is even more
crucial \cite{DGMOP}.

\item For lags greater
than one, the distributions we have defined take into account
overlapping periods. For example for lag two, one could have
defined two distributions by using non overlapping two days periods:
the distribution of the even days and the distribution of the
odd days. The number of points in these two distributions is
divided by two (for lag $L$, it would be divided by $L$)
decreasing the statistics
(in a too drastic way when $L$ is larger than four or five days for
our limited sample). This would however enable one to study
autocorrelations 
and volatility propagation.
In \cite{MDOPSM}, the authors made a precise
study of these overlapping effects, varying the overlapping intervals.
Their results were never systematically affected by these variations.
With overlapping periods, the increments are clearly not independent.
However, we will assume that they are identically distributed at a given
time scale $L$. This unconditional approach is motivated by the fact that
extreme events occur infrequently and do not seem to exhibit time dependence
\cite{Dd}. 
 
\end{enumerate}

The empirical discretized
density function $\hat f^{[m]}_L(\hat v)$ of the variations of the
interest rates $\hat v$ for every maturity $[m]$ and every
lag $L$ is defined by
\be  
\hat f^{[m]}_L(\hat v)=\frac{N^{[m]}_L(\hat v)}{N_{tot}-L}
\,.
\label{distributions}
\ee
Again, $\hat f^{[m]}_L(\hat v)$ is defined for integer values of $\hat v$,
zero outside the relevant $[\hat v_{min},\hat v_{max}]$ range and
normalized 
\be
\sum_{\hat v=\hat v_{min}}^{\hat v_{max}} \hat f^{[m]}_L(\hat v)=1
\,.
\ee 

The analogous continuous densities $f(v)$ are continuous
functions of the continuous variations $v$. We chose $\bpyear$
again as the unit of $v$. The corresponding unit for $f$ is the
inverse $\yearbp$.
They are
normalized in such a way that their integral be equal to one.
For the experimental discretized distributions
\re{distributions}, the normalizations are 
equivalent to stepwise integrals
\be
{\rm{Normalization}}_{\rm{\ data}}\equiv
\int_{v_{min}}^{v_{max}} f^{[m]}_L(v) dv
=\sum_{\hat v=\hat v_{min}}^{\hat v_{max}} 
\left( \hat f^{[m]}_L(\hat v) \right)=1
\,.
\label{normdata}
\ee
Obviously, for continuous variations $v$, 
the density
$ f^{[m]}_L(v)$ takes the constant value
$\hat f^{[m]}_L(\hat v)$ in the interval of length 
$1$ centered around 
the discretized variation $\hat v$. 
Precisely
\be
f^{[m]}_L(v)=\hat f^{[m]}_L(\hat v)
\label{contdistributions}
\ee
for
\be
\left( \hat v-\frac{1}{2} \right) 
    <v \leq 
\left( \hat v+\frac{1}{2} \right) 
\,.
\ee
Outside the $\left[ v_{min},v_{max} \right]$ range 
is $f$ taken to be zero so
that the limit in the integral can be taken to be $-\infty$ and
$+\infty$. 

The behaviour of these quantities under changes in the units is
outlined in Appendix A. 

The moments of order two of the distributions are again defined 
by a stepwise integration. To a very good approximation, since
the mean values of the distributions are essentially zero, these
second moments can be identified with the variance 
\be
{\rm{Variance}}_{\rm{\ data}}
\equiv \int_{\infty}^{\infty} 
            v^2  f^{[m]}_L(v) dv
=\sum_{\hat v=\hat v_{min}}^{\hat v_{max}} 
\left( \hat v^2 \hat f^{[m]}_L(\hat v) \right)
\,.
\label{vardata}
\ee
These variances (obtained in units $\bpyear^2$) are obviously
finite by construction. 
We however hope that the  true
variance, which is in principle
less securely computed from the data than the
normalization \re{normdata}, is correctly estimated 
by the discretized computation \re{vardata}. In
other words, we conjecture 
that the true distribution is of finite variance
contrary to some theoretical distributions like the Levy distributions
which have been
proposed in the literature 
(see \cite{Mate}, \cite{MS95},  \cite{BP})
and can lead to an infinite variance.

Later in the paper, we will discuss the variance and 
will also comment on the higher moments 
(skewness, kurtosis) of the distributions.

\subsection{General Comments}

We expect that the distributions of the variation of
interest rates in the continuous variable are smooth functions
of $v$ and hence should be approximated by 
(normalized) continuous functions. These distributions 
are essentially maximum for a variation $v=0$ and
decrease on both sides i.e. in the directions
$v\rightarrow +\infty$ 
and $v\rightarrow -\infty$. We also
expect that the distributions are, in first approximation,
symmetrical around $v=0$ and hence even functions. 
These conjectures turn out to be very well substantiated by the
experimental facts. As we will see however,
we have decided to allow for a very slight shift $s$ in the axis
of symmetry. The distribution will then enjoy the symmetry 
$(v-s)\leftrightarrow -(v-s)$. The extra
parameter $s$ will turn out to be practically zero
(less than a few $\bpyear$ and rarely different from zero at
conventional confidence level, see Table~2 and Table~3). 
In the next section we will give, in a more
precise way, our notation and our basic hypothesis.   

\section{Pad\'e Approximants}

In this section, we fix the notation which will be used for the
Pad\'e parameters. We also discuss, for the approximants we are
going to use, their normalization, their variance and their positivity.  

\subsection{Presentation of the Pad\'e Approximants}

Pad\'e Approximants $P^{[M,N]}(v)$ 
are rational functions of a continuous variable $v$ of the form 
\be
P^{[M,N]}(v)=\frac{T^M(v)}{B^N(v)}
\label{padegen}
\ee
and are indexed by the degrees $M$ and $N$ of the
polynomials $T^M$ and $B^N$ appearing in the numerator (Top) and the
denominator (Bottom) of the rational fraction $P$. 

These rational functions are used to obtain good approximations
to the continuous distributions 
\re{normdata} defined in
the introduction. Since the distributions are real, we choose the
polynomials $T$ and $B$ to depend on real parameters only.

As a general requirement, we have to limit ourselves, at the end,
to values of the parameters of the Pad\'e Approximants such that
the densities are purely positive. Indeed, even extremely
small negative probabilities in the tails would not make much
sense and should be excluded.

As we have insisted on in the introduction, we expect that the
Pad\'e Approximants (and hence the two polynomials $T$ and $B$) 
are symmetrical
under a left-right symmetry around an axis shifted at most
slightly (by $s$) from the $v=0$ axis. More precisely, we will
postulate that the two polynomials are even functions of the
variable $(v-s)$ i.e. functions of $(v-s)^2$ 
and hence that $M$
and $N$ are even numbers.

We have also commented about the fact that the
Pad\'e Approximants should be normalized 
\be
{\rm{Normalization}}_{\rm{\ Pad\acute e}}
\equiv
\int_{-\infty}^{+\infty}P^{[M,N]}(v) dv=1
\label{normpade}
\ee
and have a finite variance
\be
{\rm{Variance}}_{\rm{\ Pad\acute e}}
\equiv
\int_{-\infty}^{+\infty}P^{[M,N]}(v) (v-s)^2
dv
<\infty
\,.
\label{varpade}
\ee
The expectation (mean value) of the Pad\'e distribution is simply 
\be
{\rm{Mean\ Value}}_{\rm{\ Pad\acute e}}
\equiv
\int_{-\infty}^{+\infty}P^{[M,N]}(v) v dv
=s
\,.
\label{meanpade}
\ee
This mean value will turn out to be extremely small in absolute
value for our experimental samples.

The two general requirements \re{normpade} and 
\re{varpade} imply that the even degrees $M$ and
$N$ in \re{padegen} cannot be chosen arbitrarily. We need
\be
M+4\leq N
\,. 
\label{restriction}
\ee
  
On the other hand, though our experimental sample is reasonably
large we do not wish to introduce too many parameters in the
fit. Indeed
\begin{itemize}
\item It is known that for many classes of functions even low
values of $M$ and $N$ will give very good fits
\item We also expect that the resulting distribution should be
rather smooth. Numerical minimizations with too many parameters will push
the routines to mimic fake oscillations which are the result of
the fact that the sample is of finite size. 
These oscillations do not correspond to any intrinsic structure. 
\end{itemize}

Hence we have chosen to limit ourselves to $N=8$ and hence from
\re{restriction} to $M=4$. The relevant Pad\'e 
Appriximants belong to the
categories $[0,4],[0,6],[0,8]$, $[2,6],[2,8],[4,8]$. They can all
be written at once in the form $[4,8]$. Restricting oneself to some
parameters equated to zero will produce the other forms. If we
allow for infinite variance, the Pad\'e $[0,2],[2,4]$ and
$[4,6]$ could also be considered.

We will now write the general case in a completely explicit
form. This allows us not only to fix the notation but also to profit from
well-known integration formulas for ratios of polynomials \cite{Ryz}.

To parametrize the numerator of the Pad\'e function, 
it is convenient to introduce a complex polynomial $U^2(v)$ of
second degree
\be
U^2(v)=u_0\left( 1+i u_1 (v-s) 
               + u_2 (v-s)^2 \right)
\label{numpad1}
\ee
and write the numerator of $P^{[4,8]}$ as
\bea
T^4(v)&=&(U^2(v))^{\star} U^2(v) 
                \nonumber \\
      &=&u_0^2(1 + (u_1^2 + 2 u_2) (v-s)^2 
           + u_2^2  (v-s)^4)
                \nonumber \\
      &\equiv& n_0+n_2 (v-s)^2+n_4 (v-s)^4
\label{numpad}
\eea
which depend on three real parameters $u_i,i=0,1,2$. 
When $M$ is decreased to $2$,
$u_2=0$ and when $M$ is decreased to $0$, both $u_1=u_2=0$.    

For reasons which will be explained shortly, it is even more 
convenient to
split the denominator $B^8(v)$ of the general Pad\'e $P^{[4,8]}$
\re{padegen} in the form
of the modulus square of the complex polynomials $Q^4(v)$ of degree $4$.
\be 
Q^4(v)=1+i q_1 (v-s) 
               + q_2 (v-s)^2 
               + i q_3 (v-s)^3 
               + q_4 (v-s)^4
\label{denpad1}
\ee
where the four parameters $q_i,i=1,\ldots,4$ are real.
We then obtain the purely real eighth's degree denominator  
\bea
B^8(v)&=&(Q^4(v))^{\star} Q^4(v) 
                \nonumber \\
      &=&1 + (q_1^2 + 2 q_2) (v-s)^2 
           + (q_2^2 + 2 q_4 - 2 q_1 q_3) (v-s)^4
                \nonumber \\
      &&   + (q_3^2 + 2 q_2 q_4) (v-s)^6
           + q_4^2 (v-s)^8
                \nonumber \\
      &\equiv&1+d_2 (v-s)^2
           +d_4 (v-s)^4
           +d_6 (v-s)^6
           +d_8 (v-s)^8
\,.
\label{denpad}
\eea
It is clear that the term of zero degree 
in this denominator has been chosen to
be equal to $1$ without any loss of generality.
When $N$ is decreased to 6, $q_4=0$ and when $N$ 
is decreased to 4, $q_4=q_3=0$.

Since the form of the Pad\'e we have chosen, which is an even
real function of $(v-s)$ only, is the modulus
square of a complex quantity, the positivity of the
distribution is guaranteed whatever be the parameters chosen in
$U^{2}$ and $Q^{4}$.

\subsection{Normalization, Variance and Positivity}

The form \re{denpad} may be thought, at first sight, to be unduly
complicated. In fact, it has a very important advantage.
Indeed, the normalization and the variance can be computed
analytically \cite{Ryz}. First, the normalization \re{normpade} is
\be
{\rm{Normalization}}_{\rm{\ Pad\acute e}{[4,8]}}
     =\pi
   \frac{(q_2q_3 - q_1q_4)n_0 - q_3n_2 + q_1n_4}
        {q_1q_2q_3 - q_1^2q_4 - q_3^2}
\,.
\label{normpadean}
\ee
Imposing that this normalization be 1 decreases the number of
arbitrary parameters by one.
 
From the expression \re{normpadean} valid for the $P^{[4,8]}$,
the normalization of the Pad\'e with lower $M$ and $N$ can easily
be deduced by suitable limiting procedures. Since we will mainly
use $P^{[0,4]}$ and $P^{[0,6]}$, 
let us quote explicitly these particular cases 
\be
{\rm{Normalization}}_{\rm{\ Pad\acute e}{[0,6]}}
     =-\pi
   \frac{q_2n_0}
        {q_3-q_1q_2}
\,.
\label{normpadean60}
\ee
and
\be
{\rm{Normalization}}_{\rm{\ Pad\acute e}{[0,4]}}
     =\pi
   \frac{n_0}
        {q_1}
\label{normpadean40}
\ee
which for a normalized $P^{[0,4]}$ determines $n_0$ as
\be
n_0=\frac{q_1}{\pi}
\,.
\label{normpade40n0}
\ee

The variance \re{varpade} is 
\be
{\rm{Variance}}_{\rm{\ Pad\acute e}{[4,8]}}
    =\pi
   \frac{-q_3q_4n_0 + q_1q_4n_2 + (q_3 - q_1q_2)n_4}
        {q_4(q_1q_2q_3 - q_1^2q_4 - q_3^2)}
\,.
\label{varpadean}
\ee
For 
$P^{[0,6]}$ this reduces to
\bea
{\rm{Variance}}_{\rm{\ Pad\acute e}{[0,6]}}
    &=&\pi
   \frac{n_0}
        {q_3-q_1q_2}
    \nonumber\\
\label{varpadean60}
\eea
and for $P^{[0,4]}$ to
\bea
{\rm{Variance}}_{\rm{\ Pad\acute e}{[0,4]}}
    &=&-\pi
   \frac{n_0}
        {q_1q_2}
    \nonumber\\
&=&-\frac{1}{q_2}
\label{varpadean40}
\eea
where the last line in the equation refers to a normalized
Pad\'e ${[0,4]}$ \re{normpade40n0}.

To be complete, let us also give the expression for the kurtosis.
Obviously, in order to have a finite kurtosis, the Pad\'e have
to be restricted to $P^{[0,6]},P^{[0,8]}$ and $P^{[2,8]}$.
\be
{\rm{Kurtosis}}_{\rm{\ Pad\acute e}{[2,8]}}
    =\pi
   \frac{q_1q_4n_0 + (q_3-q_1q_2)n_2}
        {q_4(q_1q_2q_3 - q_1^2q_4 - q_3^2)\times{\rm{Variance}}^2}
\,.
\label{kurtpadean}
\ee
The excess kurtosis is given by subtracting $3$ to the kurtosis.
The kurtosis for $P^{[0,6]}$ reduces to
\bea
{\rm{Kurtosis}}_{\rm{\ Pad\acute e}{[0,6]}}
    &=&-\frac{1}{\pi}
   \frac{q_1(q_3-q_1^2)}
        {q_3 n_0}                     \nonumber\\
    &=&\frac{q_1q_2}{q_3}
\,.
\label{kurtpadean60}
\eea
The last line is for a normalized $P^{[0,6]}$.

Let us end this section by some technical points
\begin{itemize}

\item 
First, let us state again that the positivity of the distribution
follows trivially from our writing of the Pad\'e function in
terms of the modulus square of the ratio $U^{2}/Q^{4}$.

\item Let us note that, in order to obtain the values of the 
normalization, the variance and the kurtosis
starting from the formula 
given in the literature (see for example \cite{Ryz}), an
analytical continuation has been performed passing from real to
complex parameters in $Q$. A short discussion in given in
appendix C. We have checked, in each instance, that the
conditions allowing the analytical forms are satisfied.

\item Technically, Pad\'e Approximants, though they often approximate
data very well and enjoy numerous properties are not ``robust'', in
the following sense. When the powers $M$ and $N$ are changed, the
optimal values of the Pad\'e
parameters 
approximating in the best way the data 
(see \re{criteria} below) can vary widely.
In other words, the optimal value of the parameters 
for a given choice of $M$ and $N$ do not predict,
in general, in a very simple way, the values of the parameters for other
choices of $M$ and $N$. We refer the reader to the relevant
literature about this point, which has been studied widely
\cite{G}, \cite{B}.   

However, for a given value of $M$ and $N$, the parameters
entering the Pad\'e vary smoothly with respect to the lag and to
the the maturity. Hence the values for different lags and
maturities can be compared and fitted extremely well to simple forms. 
\end{itemize}

As we will show, the normalized experimental distributions
$f(v)$ (\re{distributions},
\re{contdistributions}) will be extremely well approximated by the
normalized $P^{[0,4]}$ Pad\'e Approximants.

\section{Criteria, $\chi^2$}

In order to estimate the parameters appearing in
the Pad\'e Approximants, a measure of the distance between
the experimental distribution and the Pad\'e distribution should
be chosen and minimized. 

We have decided to perform a least square fit
by defining for the (non-normalized) distributions  
$N$ (see \re{bareN},\re{normbareN}) a $\chi^2$
\be
\chi^2=\sum_{\hat v=\hat v_{min}-R}^{\hat v_{max}+R} 
     \frac{\left( P_{\rm{bare}}(\hat v)-N(\hat v)\right)^2}
          {{\sigma(\hat v)}^2}
\label{criteria}
\ee
which is discussed in the items below.

Minimization of the $\chi^2$ will be used to determine the
estimates of the
parameters $n_i$ and $q_i$ entering in the Pad\'e Approximants. 
In the formula \re{criteria}, we have, for simplicity,
suppressed all the indices referring to the lag $L$, the maturity
$[m]$ and the Pad\'e indices $[M,N]$.

Let us explain and comment on this formula as applied to our case
\begin{itemize}
\item
$P_{\rm{bare}}(\hat v)$ is the relevant Pad\'e Approximant
computed at the relevant position but normalized to the total number
of data points instead of one. The criteria itself
is a pure number and hence independent
of the unit chosen to make the computations.
If the bare $N(\hat v)$ are arbitrarily multiplied by $\mu$ both 
the bare $P$'s and
the $\sigma$'s should be multiplied by the same $\mu$.

The normalized Pad\'e Approximant $P(\hat v)$ is thus given in terms of
the bare $P_{\rm{bare}}(\hat v)$ as
\be
P(\hat v)=\frac{P_{\rm{bare}}(\hat v)}{N_{tot}-L}
\ee
in units $\yearbp$ for the variable $\hat v$ expressed in units 
$\bpyear$.

\item
The sum runs on the integer discrete integer positions 
$\hat v$ (in units $\bpyear$
which we have chosen to be our natural binning) extending at least from the
first instance $\hat v_{min}$
where $N(\hat v)$ is non zero to
the last position $\hat v_{max}$
where it is non zero. The inclusion of the integer $R\geq 0$ in \re{criteria}
allows one to take into account in the summation an extended range
of $\hat v$ where we know 
experimentally that $N(\hat v)$ is zero.
Evidently, the summation in \re{criteria} could or should 
in principle be extended to the full domain
$[-\infty,+\infty]$. Outside the initial range $[\hat v_{min},\hat v_{max}]$,  
$N(\hat v)$ is zero and $P(\hat v)$ is small but non
zero. In fact, restricting oneself to the range
$[\hat v_{min}-R,\hat v_{max}+R]$ does not change the fits very significantly. 
For large absolute values of $\hat v$, many bins have zero
observations. For this reason, the empirical characterisation of
the probability laws is sometimes performed with cumulative
distributions (see \cite{BP}). In order to keep the
implementation as simple as possible and to avoid numerical
integrations, we have decided to deal with probability densities.

\item
In order to make sense we have also to chose the ``errors''
$\sigma(\hat v)$ in the data in a suitable way. Normally, in the absence of
knowledge on the experimental errors on the $N(\hat v)$, it is customary to
chose all the $\sigma(\hat v)$ equal to the same value.
Indeed, the
``experiment'' is a one time experiment and cannot be reproduced at will.  

However, in our case, we think that
we can argue that we have some indications on the errors. If the
interest rates were purely statistically produced, we would
expect, in first approximation a random distribution around the
theoretical curve and
$\sigma(\hat v)$ for given $\hat v$ 
to be of the order of $\sqrt{N(\hat v)}$, i.e.
\be
{\rm{if}}\ N(\hat v) \neq 0\ {\rm{then}}\ \sigma(\hat v)=\sqrt{N(\hat v)}
\,.
\label{sigmadef1}
\ee
This is not exactly the case but seems to be a reasonable first guess.
Small variations about this value, due to the non randomness of
the data are obviously not important and will influence very
little the determination of the parameters obtained by
minimizing the $\chi^2$.   
When $N(\hat v)=0$, the appearance of $\sqrt{N(\hat v)}=0$ in the
denominator does not make sense. We then 
naturally chose $\sigma(\hat v)=1$,
for the $\hat v$ such that
$N(\hat v)=0$ (the minimal error on an integer number) 
\be
{\rm{if}}\ N(\hat v) = 0 \  {\rm{then}}\ \sigma(\hat v)=1 
\,.
\label{sigmadef2}
\ee
\end{itemize}

In the following section, we apply the ideas of the two last
sections to our data set.

\section{Data Analysis}

\subsection{Presentation and Statistical Description of the Data}

The raw data we have used are the
American daily spot
interest rates for constant maturities equal to one, two, 
three, five, seven, ten and thirty years 
\be
[m]=[1],\ [2],\ [3],\ [5],\ [7],\ [10],\ [30]
\label{availablemat}
\ee
between 
February 15, 1977 and August 4, 1997.
Altogether 7 rates for each of the 
\be
N_{tot}=5108
\label{Ntot}
\ee 
opening days considered. They are calculated from bond prices, 
published by the Board of Governors
of the Federal Reserve System \cite{FRS} in the United States of America
and freely available on their Web site.

In Table~1, we have given some basic statistical results which
can be obtained directly from the data : in particular the mean
, the variance, the skewness and the kurtosis for all
available maturities and for lags $L=1,5,10,15,20,25,30$. 
We have restricted ourselves to rather small time scales (up to
thirty business days). For longer periods, the sizes of the
samples become too small and the mean values cannot be neglected
safely. All the variations in interest rates exhibit the same
general behaviour : increasing variance and decreasing
leptokurticity for larger lags. The small assymmetries in the
data are obviously not reproduced by the even Pad\'e
Approximants that we have selected. 

\subsection{Generalities}

Taking into account our general consideration given above, the
precise $\chi^2$ we have chosen to minimize is
\be
\chi^2=\sum_{\hat v=\hat v_{min}-R}^{\hat v_{max}+R} 
     \frac{\left( P_{\rm{bare}}(\hat v)-N(\hat v)\right)^2}
          {N_e(\hat v)}
\label{criteriafinal}
\ee
where $N_e(\hat v)$ is defined as $N(\hat v)$ when it is non
zero and $1$ otherwise.

Let us call $f_e(\hat v)$ the ratio
\be
\hat f_e(\hat v)=\frac{N_e(\hat v)}{N_{tot}-L}
\ee
which is equal to $\hat f(\hat v)$ (see \re{distributions}) when
it is non zero but equal to $1/(N_{tot}-L)$ when $\hat f(\hat v)=0$.

The final form of the criteria is
\be
\chi^2=(N_{tot}-L)\times
          \sum_{\hat v=\hat v_{min}-R}^{\hat v_{max}+R}
          \frac{\left( P(\hat v)-\hat f(\hat v)\right)^2}
          {\hat f_e(\hat v)} 
\,.
\label{criterianormfinal}
\ee

\subsection{Fits with Pad\'e $[0,4]$. Estimation of the Parameters
for all the Values of the Lags and Maturities}
 
\subsubsection{Normalized Pad\'e $[0,4]$}

We have computed the best fit values of the parameters 
of a Pad\'e $P^{[0,4]}(\hat v)$, defined in \re{numpad},
\re{denpad} and \re{denpad1} but limited to non zero
$n_0,q_1,q_2$.
The $\chi^2$ \re{criterianormfinal} has been minimized
with a normalized Pad\'e ($n_0=q_1/\pi$)
(see \re{normpade40n0}). 

To that effect, we have used the Fortran
IMSL DUNLSF minimization programme
based on a modified Levenberg-Marquardt algorithm and a finite
difference Jacobian. 
The best values of the parameters $s$,
$q_1$ and $q_2$ (in respective units $\pcyear$, $\century$,
$\century^2$) and of their ``one parameter standard error'' 
$\Delta s$, $\Delta q_1$ and $\Delta q_2$ (in the same
units) (see appendix B) have been computed for all
values of the lag and of the maturity: altogether 210
different cases.

The results are given
in Table~2 for all maturities but for lags
$L=1,5,10,15,20,25,30$ only, to keep 
a reasonable size for the table.
The other values can be obtained from
the authors.

Some remarks can be made at this point:
\begin{enumerate}

\item 
For all the cases, including
those which are not reported in Table~2, except for 
the $[m]=[2],L=1$ case, the
hypothesis that the distribution is a Pad\'e $[0,4]$ cannot be rejected
at the $5\%$ confidence level. The exception is due to the fact
that the experimental curve has an oscillatory behaviour around
its maximum with three separate peaks. This is atypical as
compared to all the other cases. 
We conjecture that this is due to a statistical fluctuation 
and does not correspond to a
genuine and reproducible effect in the two years maturities. 

\item
The modulus of the shift parameter $\mid s\mid$ is always very small 
and can essentially be neglected. 
Indeed, it never exceeds a few $\bpyear$ and in most cases, it
is not statistically different from zero. This is a simple
reflection of the fact that interest rates themselves follow
cycles which come back to their original value.

This means that, in first approximation, the data are
symmetrical under the $(v)\leftrightarrow (-v)$ exchange, as was foreseen
and that the influence of $s$ can be safely forgotten.
Both the normalization
and variance can, well within the experimental
errors, be equated to the values obtained with $s=0$. 

\item
For all values of the maturities, the two relevant parameters
(and also $n_0$ which is by normalization a dependent parameters), 
$q_1$
and $q_2$ decrease very smoothly as a function of the lag,
including the values of the lags which are not in Table~2. 
This
is a simple reflection of the fact that as the lag becomes
higher the distribution flattens. To a very good approximation, 
these decreasing functions are straight lines in a log-log plot.
We will estimate the parameters of these straight lines in
Section~5.4.  

\item
As a function of the maturity, the parameters first decrease
then become higher showing a dip. Hence, the density function
shows, first a tendency to flatten but becomes sharper when very
long maturities are concerned. Experimental variations of the interest rates 
for the two and three
years maturities seem to be somewhat less ``smooth'' than for the
one year maturity. In turn, densities for maturities of 30 years are
remarkably stable and strongly peaked around zero variations.

\item
Even though as a function of lag and maturity together, no very
simple pattern emerges, the smoothness of the parameters in the separate
variables is a very welcome fact.

\item
We have chosen to draw, in Figure~1,
the experimental
distribution together with the one given by the best Pad\'e
for the maturity $[m]=[1]$ and the lag $L=1$ and in Figure~2 the case
$[m]=[1]$ and $L=30$. In both cases, the
agreement between the two curves is rather good. The quality of
the fit reflects the low value of the criteria equivalent to a
goodness of fit of about one. All the analogous figures for the
210 cases show the same agreement. In one case only, the
$[m]=[2]$, $L=1$ case, the agreement is worse, reflecting the
higher value of the criteria and a low goodness of fit.
The related Figure~3 shows the reason for this fact. Instead of
presenting one maximum, the experimental curve exhibits an
oscillatory pattern around zero.  
The density of variation of interest rates is strongly peaked
at $v=2$, $v=0$ and $v=-2$ instead of $0$ as is the
normal situation. 

\item
Though the Pad\'e distributions do not have the properties of the
Levy distributions, they enjoy with these distributions narrower peaks around
their maximum and longer tails as compared to the Gaussian
distributions (see below). Compared to the Levy
distributions, they have the advantage of being expressed in a very simple
rational form. As we have seen, many auxiliary computations can be
done very easily and often analytically.   

\end{enumerate}

\subsubsection{Normalized Pad\'e $[0,4]$ with Constrained Variance}

The estimations of the parameters $q_2$ allows, in principle, to
evaluate directly the variances \re{varpadean40}. It turns out
that the Pad\'e variances do not always equate the values
obtained directly from the data. 
As one can verify directly from Table~1 and Table~2, there is no
systematic pattern in the deviations~: for short maturities (1, 2
and 3 years) and large lags, the Pad\'e variances underestimate
the sample variances. The adequacy is good for intermediate
maturities. For the longest maturity (30 years), the Pad\'e
variances overestimate slightly the sample variances.

These deviations are difficult to interpret.
We have thus redone our
computations using the parameter $q_2$ constrained to 
$-1/(Sample\ Variance)$. 
Even if the minimal values obtained for the criteria are higher
than in the non-constrained cases, the value of the criteria
remains fully acceptable except again in the $[m]=[2],\ L=1$ case
which produces a too high $\chi^2$. Again, this exception is due to the
fact that the data for $[m]=[2],\ L=1$ are atypical and that the
experimental curve exhibits more than one maximum. 

The parameters $s$ are again rather small and, in most cases,
not statistically
different from zero. The values of the parameters $s$, $q_1$
and of $\chi^2$ and of their standard errors are given in Table~3. 
We have restricted ourselves again to the
lags $L=1,5,10,15,20,25,30$ to keep the table of reasonable size.
For the other lags, the parameter $q_1$ interpolates smoothly.

\subsubsection{Comparison with the Gaussian Distribution}

In order to compare the two cases corresponding to fits of
Pad\'e Approximants (with the constrained normalization and the
variance constrained or not), we have performed the same
computation for a Gaussian distribution of the form
\be
G(v)=       \frac{1}
       {\sqrt{2\pi}\sigma}
      e^{\frac{-(v-s)^2}{2\sigma^2}}
\ee
with two free parameters $\sigma$ and $s$ for every lag and 
for every maturity.
The Gaussian
distributions lead to high $\chi^2$ as compared to the Pad\'e
Approximants.
For the non-constrained estimation, the hypothesis of a normal
distribution is always rejected for lags up to five days and, in
most cases, for the $[m]=[30]$ maturity. 
The no-rejection of the Gaussian model for some large lags is
linked to a decrease in the leptokurticity with increasing time
scale. These results are coherent with similar observations on
the FX market (see \cite{MDOPSM}).
If one constrains the
standard deviation $\sigma$ to its sample value, 
the normality is strongly rejected
for all $[m],L$ cases.

As an example we have
drawn in Figure~4 the curve for $L=1$ and $[m]=[1]$ compared to
the experimental points. This is obviously worse than the
corresponding comparison for Pad\'e $[0,4]$ given in Figure~1.
The Gaussian distribution is clearly wider for low $v$ and drops
more sharply for higher $\mid v\mid$ than the data points. The
Pad\'e curve is narrower for low $v$ but drops less sharply for
higher $\mid v\mid$ and hence fits the data much better.
All the $[m],L$ cases present the same behaviour.

\subsection{Fits with a Normalized Pad\'e $[0,4]$. Variation of the 
Parameters as
a Function of the Lag for Different Maturities}

The parameters $q_1^{[m]}(L)$ and $q_2^{[m]}(L)$ at their best value 
are simple
functions of the lag $L$. More precisely, in a log-log plot, we can
approximate them by a linear model of the form
\bea
\ln(q_{1}^{[m]}(L))&=&\lambda_{1}^{[m]} \ln(\frac{L}{L_0})+\nu_{1}^{[m]}
\nonumber\\
\ln(-q_{2}^{[m]}(L))&=&\lambda_{2}^{[m]} \ln(\frac{L}{L_0})+\nu_{2}^{[m]}
\,.
\label{lanuline}
\eea
The justification of the arbitrary factor $L_0$ which will be
taken equal to $15$ is given in
Appendix B where a precise discussion of the 
related standard errors is given.

The $q_{i}^{[m]}(L)$ are expressed in some units (see Appendix A). 
If these units are changed,
$\lambda_{i}^{[m]}$ does not change while $\nu_{i}^{[m]}$ picks
up, obviously, an additive constants (the logarithm of the ratio between the
two units). 

For the estimation of $\lambda_i^{[m]}$ and $\sigma_i^{[m]}$,
we have to define new $\chi^2$ ($C_i^{[m]}$) for
every $q_i$ and every maturity $[m]$. This takes into
account the errors we have obtained on the $q_i$. They are
\bea
C_1^{[m]}&=&\sum_{L=1}^{30}
       \frac{\left(e^{\nu_{1}^{[m]}}(\frac{L}{L_0})^{\lambda_{1}^{[m]}}
                       -q_{1}^{[m]}(L)\right)^2}
            {(\Delta q_{1}^{[m]}(L))^2}
   \nonumber\\
C_2^{[m]}&=&\sum_{L=1}^{30}
       \frac{\left(e^{\nu_{2}^{[m]}}(\frac{L}{L_0})^{\lambda_{2}^{[m]}}
                       +q_{2}^{[m]}(L)\right)^2}
            {(\Delta q_{2}^{[m]}(L))^2}
\,.
\label{criterialambdanu}
\eea

In Table~4, the values of the best $\lambda_{i}^{[m]}$ and
$\nu_{i}^{[m]}$ minimizing $C_i^{[m]}$ and their 
one standard error $\Delta \lambda_{i}^{[m]}$ 
and $\Delta \nu_{i}^{[m]}$ are given for all maturities. 
In Figure~5 and Figure~6, the straight lines \re{lanuline} are plotted
for $[m]=[1]$ and compared to the data points. The fit is clearly
excellent for $\ln(q_1^{[1]})$
and somewhat worse for $\ln(-q_2^{[1]})$.

The parameters $\lambda_2$ are related to the so-called
``scaling laws'' for the time dependence of the standard deviations
\be
\sigma^{[m]}(L)=k^{[m]} L^{E^{[m]}}
\,.
\ee
This behaviour is reported for most of the financial time series
(see \cite{MS97}, \cite{GBPTD}, \cite{MDOPSM}) with an exponent
typically close to but larger than $0.5$, the value which should
be observed for a Gaussian process. 
With our notations $E^{[m]}=-\lambda_2^{[m]}/2$. The absolute
value of $\lambda_2$, statistically greater than 1, lead to
scaling exponents ranging from $0.52$ to $0.54$.
 
\subsection{Discussion of Fits with Pad\'e Approximants with
Indices higher than $[0,4]$}

Since the Pad\'e Approximants with indices $[0,4]$ lead to such
good fits to the data, it is tempting to see if these fits can
even be made better by using higher indices. We have tried to do
so and have come to the conclusion that $[0,4]$ is close to the
optimum. 

\begin{enumerate}
\item
First, the Pad\'e Approximants $[M,N]$, when the condition
of the existence of a finite variance is not satisfied \re{varpade}, 
do not seem to be very probable. 
Those with $M$=$N$ clearly are not convenient as they cannot
even be normalized.
    
\item
When the condition of finite variance is satisfied \re{varpade},
Pad\'e with indices higher that $[0,4]$ obviously produce
somewhat better fits and hence lower values for the $\chi^2$.
But this statement has to be qualified by considering in turn
the different possibilities.
\begin{enumerate}
\item
When $[0,6]$ is used, the parameter $q_3$, new with respect to
our first choice, is significantly different from zero in few
cases only.
The $\chi^2$ is not improved in most cases.
This means that the $[0,4]$ approximation is already able to
cope with the decrease of the experimental density function. A
$v^{-4}$ is sufficient to achieve it. 
Our careful, but, may be, slightly prejudiced
analysis of the data even seems to point to the fact that the
$v^{-4}$ decrease is the fastest which can be allowed by the data.
Indeed, the fit to the variance ($-1/q_2$) coming out of the
minimization already tends to come out too small as compared to the
raw variance of the data, for maturities up to three years. 
Remember that this raw variance
depends crucially on the end points of the experimental
distribution and hence is not determined very precisely.  
For $P^{[0,6]}$, the moment of fourth degree (numerator of the
kurtosis) then becomes finite which may be interesting
theoretically though to repeat, we believe that the data point in
the infinite kurtosis direction. The same findings were achieved
on the FX rates in \cite{DGMOP}.

\item
We have also tried to fit the experimental data with the other Pad\'e's
($[0,8]$, $[2,8]$, $[4,8]$ and $[2,6]$).
In the majority of the cases, no significant improvement of the
$\chi^2$ is obtained. 
In some cases, the minimization algorithm fails $[2,6]$.
In a few cases
our minimizations lead to a smaller value for the criteria. But,
the price to pay is that the Pad\'e curve starts to try to fit small
oscillations in the experimental curves $[0,8]$, $[2,8]$ and $[4,8]$. 
These oscillations are
obviously due
to the finiteness of the sample and do not very likely correspond to
intrinsic structures.
Moreover, the minimum obtained are usually very sensitive to
the starting values and the minimal parameters do not exhibit
the stability they show in  the $[0,4]$ case.
\end{enumerate}
\end{enumerate}

\section{Conclusions}

In this paper, we have presented and discussed phenomenological
fits of the daily variations of 
the term structure of interest rates (taking as an
example those published by the Board of Governors
of the Federal Reserve System \cite{FRS}). 
We have shown that simple Pad\'e Approximants, which are
theoretically well suited for approximating in a rather smooth
way many classes of functions, fit the experimental results
amazingly well. 

The best values of the parameters defining the Pad\'e Approximants
vary rather smoothly as a function of the lag and of the
maturity. In particular, as a function of the lag, in a log-log
plot the parameters are very well represented by straight lines.
This can be related to scaling laws reported for other financial
time series.

We have also shown that, the simplest Pad\'e Approximant $P^{[0,4]}$,
which depends essentially on only two meaningful parameters, has
enough richness to represent the data faithfully and smoothly.
The extension to higher Pad\'e Approximants (for example to
$P^{[0,6]}$) is by no means necessary. The Pad\'e
Approximants have then a
tendency to mimic small oscillations in the data. These
oscillations are clearly the results of the finiteness of the
sample and do not seem to be related to any ``real'' underlying
structure. 

Since Pad\'e Approximants are simple rational functions, they can
be easily written, tabulated, and, once the best parameters have
been obtained, used for further applications.
We hope to come back to this idea in the near future and show
how these explicit results can be used for estimating risks
related to extreme events.
 
\appendix
\section{Dimensions of the Pad\'e Parameters}

A short appendix about units and dimensions is useful here. 
Interest rates have
the dimension of the inverse of time, $\timedim^{-1}$,
are usually given in $\pcyear$ with exactly two meaningful
digits and hence are expressed in $\bpyear$ as an integer number. 
\begin{itemize}
\item
This makes
$\bpyear=10^{-4}/\year$ a 
very natural unit to use both for the interest rates but even more
for the variation in the interest rates. It gives a natural binning
for the discretized distributions. 
The inverse unit of $\bpyear$ is $\yearbp$ which is the time scale
of $10^4$ years. This is the unit we have consistently used in the
main body of the paper.
\item
In the tables and in the figures,
the different parameters as well as the
means and the variances are expressed as
powers of the
more conventional unit $\pcyear$ and of its inverse $\yearpc$ which
is a $\century$. They are units of well adapted sizes.
\item 
In the following table, we quote the dimension $[[x]]$ of the
quantities $x$ we have been using, the units $[x]$ we have been
using in the main part of the text and the units we have used in
the tables.
\end{itemize}
\vskip 0.5 true cm
\centerline{Table of Dimensions and Units}
\vskip 1 true cm
\mbox{
\hskip 1 true cm
\begin{tabular}
{| c | l | l | l | l |}
\hline
	&		&[unit]	used	&[unit] 	& \\
variable&[[dimension]]	&in the body	&used in the 	&restriction \\
	&		&of the paper	&tables 	&\\
\hline
\hline
$v$ 		&\td$^{-1}$    	&\bpy 		&\pcy 		&\\
\cline{1-4}
$s$ 		&\td$^{-1}$    	&\bpy 		&\pcy 		&\\	
\cline{1-4}
$f$ 		&\td     	&\ybp		&\ypc		&\\
\cline{1-4}
$P$ 		&\td     	&\ybp		&\ypc		&\\
\cline{1-4}
$n_i$ 		&\td$^{1+i}$   	&\ybp$^{i+1}$	&\ypc$^{i+1}$	&\\
\cline{1-4}
$u_0$ 		&\td\half 	&\ybp\half	&\ypc\half 	&\\
\hline
$u_i$ 		&\td$^{i}$	&\ybp$^i$	&\ypc$^i$	&for $i=1,2$   \\
\hline
$q_i$ 		&\td$^{i}$    	&\ybp$^i$ 	&\ypc$^i$ 	&\\
\cline{1-4}
$d_i$ 		&\td$^{i}$    	&\ybp$^i$ 	&\ypc$^i$	&\\
\cline{1-4}
Mean 		&\td$^{-1}$	&\bpy	 	&\pcy		&\\
\cline{1-4}
Variance	&\td$^{-2}$	&\bpy$^2$ 	&\pcy$^2$	&\\
\cline{1-4}
Skewness	&\td$^0$	&		&		&\\
\cline{1-2}
Kurtosis 	&\td$^0$	&		&		&\\
\cline{1-2}
$\chi^2$ 	&\td$^0$	&		&		&\\
\hline\hline
\end{tabular}
}
\vskip 1 true cm
\begin{itemize}
\item
It is easy to pass from variables expressed in one unit to the
variables expressed in the other unit by using the correspondence
\bea
\bpyear &=& \frac{\pcyear}{100}
      \nonumber\\
\yearbp&=&100\times \yearpc=100\times{\rm{century}}
\,.
\label{unitchange}
\eea

\item
For example, the value of $q_i$ expressed in units $\yearbp^i$
becomes $100^i\times q_i$ in the more customary unit $\yearpc^i$
while $n_i$ becomes $100^{i+1}\times n_i$. The
Pad\'e density $P(v,n_i,q_i)$ (where we have written explicitly
the parameters \re{numpad},\re{denpad1})
in units $\yearbp$ for $v$ expressed in $\bpyear$
is related to the density $P^{new}(v)$ where $P^{new}$ is
expressed in $\century$ in terms of $v$ expressed in $\pcyear$ by
\be
P^{new}(v)=P(v;100^{i+1}n_i,100^{i}q_i)
\,.
\label{Punit change}
\ee
\end{itemize}

\section{Discussion of the Errors}  

A short discussion on the determination of the errors is
worthwhile giving. The $\chi^2$ 
(\re{criterianormfinal} or \re{criterialambdanu}), 
depend on
the Pad\'e or on the $\lambda,\nu$ parameters. 
Let us call these parameters generically
$p=\{p_1,p_2,\ldots\}$. The best values of the parameters 
$p^{best}$ are obtained by minimizing the
criteria $\chi^2(p^{best})=\chi^2_{min}$. 
The confidence regions in one parameter with confidence level
$p=68.3\%$ are obtained for the contour 
\be
\Delta \chi^2\equiv  \chi^2(p)-\chi^2_{min}=1
\,.
\label{standarddeviation1}
\ee
For errors on two parameters jointly, the contour is given by
\be
\Delta \chi^2= 2.3
\,.
\label{standarddeviation23}
\ee

From the computation 
of the matrix $H$ of the second derivatives 
(the Hessian) of the Criteria $\chi^2$
at its minimum and from the Taylor expansion
 of $\chi^2(p)$
(to second order in the parameters) one finds
\be
\Delta \chi^2= \frac{1}{2} H_{ij}(p_i-p^{best}_i)(p_j-p^{best}_j)
\,.
\label{critexpansion}      
\ee
\begin{itemize}
\item
If this ellipsoid has its principal axis reasonably aligned
along the parameter axis 
(which is the case for the parameters we have
chosen as can be seen on Figures~7, 8 and 9) one can define the
``one parameter standard error'' $\Delta p_i$, for a given parameter $p_i$
as
\be
\Delta p_i\equiv\mid p_i-p^{best}_i\mid=
        \sqrt{\frac{2}{H_{ii}}}
\,.
\label{oneparerror}
\ee
Notice that the classical interpretation of standard errors is
only valid if the measurements errors are normally distributed.
If they are not, a suitable contour of constant $\chi^2$
(\re{standarddeviation1}, \re{standarddeviation23}) should
be used as boundary of the confidence region (see \cite{NR}) for
a more precise discussion). 

\item
If this ellipsoid does not have its principal axis reasonably aligned
along the axis, the ``one parameter error'' is obtained from
the maximal $M_i$ and minimal $m_i$ values 
of the fixed parameter $p_i$ in
the region $\Delta \chi^2\leq 1$ as
\be
\Delta p_i=\frac{M_i-m_i}{2}
\,.
\ee

\end{itemize}

It should be remarked that for \re{criterialambdanu}, the
(obvious)
value $L_0=15$ is close to the optimal value which makes the
principal axis of the ellipse parallel to the parameter axis.

As examples, the curves corresponding to $\Delta \chi^2=1$ and
$\Delta \chi^2=2.3$ have been drawn
for the parameters $q_1,q_2$ in the $[m]=[1],\ L=1$
case in Figure~7, for the parameters $\lambda_1$ and$\nu_1$
for $[m]=[1]$ in Figure~8 and for the parameters $\lambda_2$ and$\nu_2$
for $[m]=[1]$ in Figure~9 .   

\section{Analyticity of the Denominator}

The Normalization \re{normpadean}, the Variance \re{varpadean} 
and the Kurtosis
\re{kurtpadean} have been computed analytically from the form we have
given for the denominator of the Pad\'e Approximant \re{denpad1}. There
is a restriction for these formulas to be valid, namely that
the poles of $Q^n(v)$ be all situated in, say, the upper-half
plane of the complexified variable $v$. All the poles of
$Q^{n\star}(v)$ are then situated in the lower half-plane. 
If the poles have complex values $v_j=s+a_j+ib_j,\ j=1,\ldots,4$, 
with $b_j >0$, $Q^n(v)$
is written
\be
Q^n(v)=\prod_{j=1}^{n} (1+i\frac{v-s}{v_j})
\,.
\ee

Expanding the products and taking account of the positivity of
the $b_j$, 
it is easy to see this implies for $Q^3$ the necessary conditions
are

\mbox{
\hskip 5 true cm
\begin{tabular}
{l l}
$q_1>0,$    
     &$q_2<0$   \\
$q_3\leq 0,$   
     &       $q_3-q_1q_2\geq 0$ \,.\\
\label{necessarypos3}
\end{tabular}
}

We expect that these conditions, for $Q^4$, are supplemented
by the conditions

\mbox{
\hskip 5 true cm
\begin{tabular}
{l l
}
$q_4\geq 0,$ 
     & $q_2q_3-q_1q_4 \geq 0$   \\    
     &$q_1q_2q_3-q_1^2q_4-q_3^2\geq 0$   \\
\label{necessarypos4}
\end{tabular}
}

Moreover, written in terms of the poles, we have verified
that all the analytical forms of the Normalization, of the
Variance and the Kurtosis lead to positive values and hence are
fully meaningful.

\goodbreak

\vfill\newpage
\centerline{Table Captions}
\begin{itemize}

\item Table~1

\noindent Table of Univariate Statistics for the daily changes
of the American Spot Interest Rates (expressed 
in the relevant $\pcyear$ unit)
between February 15, 1977
and August 4, 1997. The statistics are given for all available
maturities $[m]=[1],[2],[3]$,$[5]$,$[7]$,$[10]$,$[30]$ and for the 
representative subset
of lags $L=1,5,10,15,20,25$ and $L=30$. The quantities are
expressed in the relevant $\pcyear$ unit. See appendix A. 

\item Table~2

\noindent The optimal values of the parameters $q_1$,
$q_2$ and $s$ 
appearing in the denominator of the Pad\'e $P^{[0,4]}$ together with their
``errors'' \dqu, \dqd\  and \ds \ (for the units chosen, see appendix
A, for the interpretation of the errors, see appendix B). 
The parameters $n_0$ are constrained by the normalization.
These
optimal parameters are given for all available maturities
$[m]=[1],[2],[3],[5],[7],[10],[30]$ and for 
the representative subset
of lags $L=1,5,10,15,20,25$ and $L=30$. The parameters for the other values of
the lag have been computed but are not given to limit the size
of the table. They vary smoothly and have been
taken into account to estimate the parameters $\lambda_i$ and
$\nu_i$ appearing in Table~4.
The minimal value of the Criteria 
\C \ (a pure number) is also given. For 
all the cases, including
those which are not reported in the table, except for 
the $[m]=[2],\ L=1$ case
whose \C\  is marked with a $\star$ in the table, the
hypothesis that the distribution is a Pad\'e $[0,4]$ cannot be rejected
at the $5\%$ confidence level. 

\item Table~3

\noindent The optimal values of the parameters $q_1$ and $s$ 
appearing in the denominator of the Pad\'e $P^{[0,4]}$ together with their
errors \dqu, \ds\  (for the units chosen, see appendix A, for the
interpretation of the errors, see appendix B).  
In this table, the values of $n_0$ are constrained by the
normalization and the values of $q_2$ by the
experimental variance.
These
parameters are given for all available maturities
$[m]=[1],[2],[3],[5],[7],[10]$,$[30]$ and for 
the representative subset
of lags $L=1,5,10,15,20,25,30$. The parameters for the other values of
the lag have been computed but are not given to limit the size
of the table. 
They vary smoothly.
The minimal value of the Criteria 
\C \ (a pure number) is also given. For 
all the cases, including
those which are not reported in the table, except for 
the $[m]=[2],\ L=1$ case
whose \C\  is marked with a $\star$ in the table, the
hypothesis that the distribution is a Pad\'e $[0,4]$ cannot be rejected
at the $5\%$ confidence level. 

\item Table~4

\noindent The optimal values of the \laqu, \nuqu, \laqd, \nuqd 
\ parameters and their standard errors \dlaqu, \dnuqu, \dlaqd, \dnuqd
\ as functions of the maturity 
(for the interpretation of the errors, see appendix B).

\end{itemize}
\vfill\newpage

\centerline{Figure Captions}
\begin{itemize}

\item Figure~1a

\noindent The experimental
distribution compared with the one given by the best Pad\'e $[0,4]$
for maturity $[m]=[1]$ and lag $L=1$.

\item Figure~1b

\noindent The difference between the experimental 
distribution and the
distribution obtained by the best Pad\'e $[0,4]$
for maturity $[m]=[1]$ and lag $L=1$. Comparing this figure to
Figure~1a, one sees that, in absolute value, the deviation
between the data and the
approximation is larger in the central part of the distribution
but with no systematic bias. 

\item Figure~1c 

\noindent The ratio $R$ between the experimental density and the
best Pad\'e $[0,4]$ density for maturity $[m]=[1]$ and lag $L=1$.
This figure zooms the tails as compared to the central values.
One sees that, in relative size, the central part is very well
reproduced. For absolute values of $v$ higher than 0.4,
the scarcity of
the data is reflected by the appearance in the plot of
the horizontal axis and of  
curves rising as $v^4$. 
These correspond to the rare events
$N(\hat v)=0,1,2,3$ (see \re{bareN}) in the tails of the 
distribution of variations. In this region, the Pad\'e
distribution averages the data.   

\noindent 

\item Figure~2

\noindent The experimental
distribution compared with the one given by the best Pad\'e
$[0,4]$  for the maturity $[m]=[1]$ and the lag $L=30$.

\item Figure~3

\noindent The exceptional case of the experimental
distribution compared with the one given by the best Pad\'e
$[0,4]$ 
for the maturity $[m]=[2]$ and the lag $L=1$. The experimental
points exhibit three very large oscillations 
in the region around $v=0$ showed in the figure.
We have highlighted this fact by arbitrarily drawing the dotted line. 
These oscillations may be of statistical nature as they do not
show up in any of the 209 other cases that we have studied.

\item Figure~4a

\noindent The experimental distribution compared to the Gaussian
distribution for the maturity $[m]=[1]$ and the lag $L=1$.
The experimental distribution is more strongly peaked than the
Gaussian distribution which itself has too small tails.

\item Figure~4b

\noindent The difference between the experimental distribution
and the Gaussian
distribution for the maturity $[m]=[1]$ and the lag $L=1$. This
figure is worth comparing with Figure~1b.
One sees that, in absolute value, the deviation
between the data and the
approximation is large in the central part of the distribution
but also shows a systematic bias in the region  
$0.2\leq \mid v\mid \leq 0.4$. 

\item Figure~4c

\noindent The ratio $R$ between the experimental density and the
best Gaussian density for maturity $[m]=[1]$ and lag $L=1$.
This figure zooms the tails as compared to the central values.
One sees that, in relative size, 
neither the central part nor the tails are very well
reproduced. For absolute values of $v$ higher than 0.2,
$R$ is seen to shoot up extremely fast. Hence, the tails are very badly
fitted by a Gaussian.

\item Figure~5

The straight line $\ln(q_1)$ (see \re{lanuline}) as a function
of $\ln (L/L_0)$ compared with the data points for $[m]=[1]$. 
The parameters are obtained from
the best fit with a normalized Pad\'e $[0.4]$.

\item Figure~6

The straight line $\ln(-q_2)$ (see \re{lanuline}) 
as a function of 
$\ln (L/L_0)$ compared with the data points for $[m]=[1]$.

\item Figure~7

The ellipses corresponding to $\Delta \chi^2=1$ and $\Delta
\chi^2=2.3$ for the errors in the parameters $q_1,q_2$ appearing
in the $[m]=[1],L=1$ case. The estimates of $q_1$ and $q_2$ are found
in Table~2.

\item Figure~8

The ellipses corresponding to $\Delta \chi^2=1$ and 
$\Delta \chi^2=2.3$ for the errors 
in the parameters $\lambda_1,\nu_1$ appearing
in the linear model of $\ln(q_1)$ as a function of $\ln(L/L_0)$ in
the $[m]=[1]$ case. The estimates of $\lambda_1$ and $\nu_1$ are
found in Table~4. 

\item Figure~9

The ellipses corresponding to $\Delta \chi^2=1$ and $\Delta
\chi^2=2.3$ for the errors in the parameters $\lambda_2,\nu_2$ appearing
in the linear model of $\ln(-q_2)$ as a function of $\ln(L/L_0)$ in
the $[m]=[1]$ case. The estimates of $\lambda_2$ and $\nu_2$ are
found in Table~4. 

\end{itemize}
\vfill\newpage

\centerline{Table~1}
\centerline{Table of Univariate Statistics}
\vskip 1 true cm
\mbox{
\small{
\begin{tabular}
{| c | c | r | r | r | r | r | r | r |}
\hline
	&L & 1 & 5 & 10 & 15 & 20 & 25 & 30 \\
\hline
\hline
	&Mean & 0.0000 & 0.0001 & 0.0000 & 0.0001 
	& 0.0001 & 0.0002 & 0.0004 \\
	&Variance & 0.0143 & 0.0860 & 0.1966 & 0.3326 
	& 0.4836 & 0.6446 & 0.8136 \\
$[$m]=$[$1]	&Skewness & -0.1541 & -0.8223 & -1.0587 & -1.0710 
	& -1.1273 & -1.1413 & -1.1594 \\
	&Kurtosis & 14.5664 & 10.7279 & 10.4445 & 10.9208 
	& 10.9326 & 10.9088 & 10.8980 \\
	&$v_{min}$ & -1.0800 & -2.2700 & -3.0600 & -4.0700 
	& -5.4500 & -6.3100 & -6.9100 \\
	&$v_{max}$ & 1.1000 & 2.0600 & 2.4700 & 3.1400 
	& 3.2000 & 3.7200 & 4.0100 \\
\hline
	&Mean & 0.0000 & -0.0002 & -0.0005 & -0.0007 
	& -0.0009 & -0.0009 & -0.0010 \\
	&Variance & 0.0113 & 0.0702 & 0.1594 & 0.2663 
	& 0.3836 & 0.5109 & 0.6456 \\
$[$m]=$[$2]	&Skewness & -0.3648 & -0.7398 & -0.8722 & -0.8171 
	& -0.8547 & -0.8600 & -0.8741 \\
	&Kurtosis & 12.4522 & 9.7384 & 9.1389 & 9.2421 
	& 8.8631 & 8.5522 & 8.3537 \\
	&$v_{min}$ & -0.8400 & -2.0800 & -2.7900 & -3.3700 
	& -4.4900 & -5.2300 & -5.9600 \\
	&$v_{max}$ & 0.8900 & 1.9900 & 2.6000 & 2.8700 
	& 3.0900 & 3.3900 & 3.7100 \\
\hline
	&Mean & -0.0001 & -0.0004 & -0.0010 & -0.0015 
	& -0.0019 & -0.0022 & -0.0025 \\
	&Variance & 0.0102 & 0.0622 & 0.1374 & 0.2256 
	& 0.3225 & 0.4285 & 0.5421 \\
$[$m]=$[$3]	&Skewness & -0.1628 & -0.4432 & -0.5977 & -0.5000 
	& -0.5254 & -0.5487 & -0.5928 \\
	&Kurtosis & 10.4136 & 7.7748 & 7.6880 & 7.4571 
	& 6.7799 & 6.4504 & 6.3316 \\
	&$v_{min}$ & -0.7900 & -1.5700 & -2.8300 & -3.0500 
	& -3.5800 & -4.3200 & -5.1200 \\
	&$v_{max}$ & 0.9200 & 1.9900 & 2.6000 & 3.0400 
	& 3.3200 & 3.6000 & 3.6500 \\
\hline
	&Mean & -0.0001 & -0.0007 & -0.0016 & -0.0024 
	& -0.0032 & -0.0038 & -0.0044 \\
	&Variance & 0.0091 & 0.0544 & 0.1182 & 0.1901 
	& 0.2661 & 0.3491 & 0.4409 \\
$[$m]=$[$5]	&Skewness & -0.3064 & -0.3938 & -0.4482 & -0.3617 
	& -0.3448 & -0.3109 & -0.3025 \\
	&Kurtosis & 8.7216 & 6.2382 & 5.9828 & 5.9650 
	& 5.5396 & 4.8740 & 4.4440 \\
	&$v_{min}$ & -0.7700 & -1.5400 & -2.3900 & -2.5800 
	& -3.0700 & -3.6000 & -4.3200 \\
	&$v_{max}$ & 0.7200 & 1.6400 & 2.3600 & 2.8200 
	& 3.0800 & 3.3800 & 3.5100 \\
\hline
	&Mean & -0.0002 & -0.0010 & -0.0021 & -0.0031 
	& -0.0040 & -0.0048 & -0.0056 \\
	&Variance & 0.0085 & 0.0491 & 0.1044 & 0.1651 
	& 0.2287 & 0.2981 & 0.3749 \\
$[$m]=$[$7]	&Skewness & -0.3066 & -0.3907 & -0.4981 & -0.3602 
	& -0.3019 & -0.2683 & -0.2637 \\
	&Kurtosis & 8.1780 & 5.3213 & 5.4510 & 5.2149 
	& 4.6201 & 3.9410 & 3.5034 \\
	&$v_{min}$ & -0.7800 & -1.3600 & -2.4000 & -2.5500 
	& -2.9400 & -3.1100 & -3.6200 \\
	&$v_{max}$ & 0.7000 & 1.5300 & 2.0300 & 2.6000 
	& 2.8000 & 3.1100 & 3.2600 \\
\hline
	&Mean & -0.0002 & -0.0012 & -0.0025 & -0.0037 
	& -0.0049 & -0.0060 & -0.0070 \\
	&Variance & 0.0076 & 0.0440 & 0.0928 & 0.1449 
	& 0.1984 & 0.2574 & 0.3224 \\
$[$m]=$[$10]&Skewness & -0.2817 & -0.5431 & -0.6719 & -0.4999 
	& -0.3918 & -0.3246 & -0.3022 \\
	&Kurtosis & 6.9688 & 5.4452 & 5.7715 & 5.1414 
	& 4.1992 & 3.4414 & 3.0903 \\
	&$v_{min}$ & -0.7500 & -1.3600 & -2.3500 & -2.6000 
	& -2.6200 & -2.8400 & -3.3700 \\
	&$v_{max}$ & 0.6500 & 1.3600 & 1.8500 & 2.3600 
	& 2.5500 & 2.8300 & 2.9700 \\
\hline
	&Mean & -0.0002 & -0.0013 & -0.0027 & -0.0039 
	& -0.0052 & -0.0062 & -0.0073 \\
	&Variance & 0.0062 & 0.0349 & 0.0723 & 0.1114 
	& 0.1518 & 0.1972 & 0.2475 \\
$[$m]=$[$30]&Skewness & -0.2245 & -0.4463 & -0.4792 & -0.3334 
	& -0.2077 & -0.1437 & -0.1078 \\
	&Kurtosis & 6.8619 & 4.5314 & 4.0574 & 3.7200 
	& 3.1661 & 2.6781 & 2.3830 \\
	&$v_{min}$ & -0.7600 & -1.3100 & -1.8800 & -2.1600 
	& -2.3800 & -2.2000 & -2.4800 \\
	&$v_{max}$ & 0.5000 & 0.9500 & 1.3800 & 1.6500 
	& 2.0800 & 2.3700 & 2.4900 \\
\hline\hline
\end{tabular}
}
}

\vfill\newpage

\centerline{Table~2}
\centerline{Parameters for the Normalized Pad\'e Approximant $[0,4]$}
\vskip 0.5 true cm
\mbox{
\footnotesize{
\begin{tabular}
{| c  | c  | c  | c  | c  | c  | c  | c  | c  | }
\hline
	&		&\LL=1		&\LL=5		&\LL=10		
	&\LL=15		&\LL=20		&\LL=25		&\LL=30\\
\hline\hline
	& \C		&140.59		&310.90		&446.62		
	&508.29		&668.08		&660.21		&665.98\\
	& \qu		&25.253		&8.633		&5.655		
	&4.474		&3.648		&3.162		&2.885\\
$[$m]=$[$1] & \dqu		&0.518		&0.171		&0.110		
	&0.085		&0.067		&0.062		&0.056\\
	& \qd		&-70.861	&-16.802	&-9.145		
	&-5.472		&-4.053		&-2.867		&-2.125\\
	& \dqd		&4.987		&0.814		&0.369		
	&0.257		&0.177		&0.130		&0.110\\
	&\ss		&-0.00071	&0.00549	&0.00968	
	&0.01143	&0.02594	&0.01579	&0.01005\\
	&\ds		&0.00078	&0.00227	&0.00331	
	&0.00419	&0.00495	&0.00584	&0.00654\\
\hline
	& \C		&367.28$^*$	&275.87		&361.33		
	&453.41		&527.96		&521.65		&534.96\\
	& \qu		&24.073		&8.022		&5.139		
	&3.940		&3.161		&2.749		&2.511\\
$[$m]=$[$2]	& \dqu		&0.475		&0.163		&0.100		
	&0.074		&0.058		&0.052		&0.047\\
	& \qd		&-84.096	&-17.733	&-9.349		
	&-5.917		&-4.286		&-3.023		&-2.204\\
	& \dqd		&4.744		&0.683		&0.338		
	&0.227		&0.151		&0.109		&0.086\\
	&\ss		&-0.00009	&0.00506	&0.01052	
	&0.01278	&0.02027	&0.01102	&-0.00135\\
	&\ds		&0.00081	&0.00240	&0.00349	
	&0.00434	&0.00512	&0.00616	&0.00709\\
\hline
	& \C		&123.82		&272.21		&324.87		
	&372.72		&504.58		&513.08		&514.94\\
	& \qu		&24.291		&7.850		&5.099		
	&3.8345		&3.158		&2.765		&2.528\\
$[$m]=$[$3]	& \dqu		&0.497		&0.161		&0.100		
	&0.073		&0.057		&0.051		&0.047\\
	& \qd		&-89.184	&-18.659	&-8.811		
	&-5.438		&-4.091		&-2.894		&-2.132\\
	& \dqd		&4.989		&0.674		&0.309		
	&0.206		&0.153		&0.107		&0.089\\
	&\ss		&0.00016	&0.00512	&0.00563	
	&0.00477	&0.01984	&0.01253	&0.001\\
	&\ds		&0.00083	&0.00241	&0.00359	
	&0.00458	&0.00523	&0.00625	&0.00711\\
\hline
	& \C		&130.61		&205.42		&280.57		
	&330.08		&428.37		&500.15		&501.82\\
	& \qu		&24.225		&7.976		&5.295		
	&4.099		&3.387		&2.988		&2.758\\
$[$m]=$[$5]	& \dqu		&0.510		&0.159		&0.103		
	&0.079		&0.063		&0.056		&0.051\\
	& \qd		&-96.702	&-18.222	&-9.425		
	&-5.844		&-4.391		&-3.092		&-2.313\\
	& \dqd		&4.700		&0.724		&0.356		
	&0.233		&0.166		&0.120		&0.103\\
	&\ss		&0.00006	&0.00206	&0.00659	
	&0.00236	&0.00124	&-0.00026	&-0.00424\\
	&\ds		&0.00082	&0.00240	&0.00346	
	&0.00439	&0.00509	&0.00599	&0.00669\\
\hline
	& \C		&124.42		&170.78		&260.73		
	&335.07		&339.22		&367.99		&516.64\\
	& \qu		&23.745		&8.038		&5.602		
	&4.355		&3.546		&3.103		&2.945\\
$[$m]=$[$7]	& \dqu		&0.492		&0.161		&0.111		
	&0.085		&0.066		&0.059		&0.053\\
	& \qd		&-100.715	&-19.25		&-9.537		
	&-6.209		&-4.393		&-3.324		&-2.860\\
	& \dqd		&4.739		&0.751		&0.375		
	&0.244		&0.174		&0.131		&0.132\\
	&\ss		&-0.0004	&0.00177	&0.00447	
	&0.00245	&-0.00041	&0.00133	&-0.00934\\
	&\ds		&0.00084	&0.00238	&0.00338	
	&0.00423	&0.00506	&0.00582	&0.00607\\
\hline
	& \C		&137.93		&175.08		&229.76		
	&319.02		&377.43		&321.47		&499.17\\
	& \qu		&24.505		&8.527		&5.815		
	&4.585		&3.824		&3.291		&3.126\\
$[$m]=$[$10]& \dqu		&0.505		&0.169		&0.113		
	&0.091		&0.073		&0.064		&0.058\\
	& \qd		&-109.545	&-21.157	&-10.172	
	&-7.088		&-5.026		&-3.594		&-3.114\\
	& \dqd		&5.035		&0.825		&0.405		
	&0.271		&0.193		&0.143		&0.139\\
	&\ss		&-0.00074	&0.00204	&0.00706	
	&0.00224	&0.0033		&0.00055	&0.00475\\
	&\ds		&0.00081	&0.00225	&0.00326	
	&0.00399	&0.00474	&0.0056		&0.00581\\
\hline
	& \C		&102.78		&166.29		&238.66		
	&280.98		&337.20		&299.56		&464.49\\
	& \qu		&27.439		&9.848		&6.878		
	&5.481		&4.548		&3.912		&3.656\\
$[$m]=$[$30]& \dqu		&0.579		&0.196		&0.135		
	&0.107		&0.087		&0.075		&0.070\\
	& \qd		&-128.242	&-25.041	&-12.123	
	&-7.921		&-5.759		&-4.234		&-3.398\\
	& \dqd		&5.661		&1.028		&0.518		
	&0.340		&0.257		&0.189		&0.167\\
	&\ss		&-0.00051	&0.00134	&0.00463	
	&0.00145	&0.0018		&0.00408	&-0.00445\\
	&\ds		&0.00073	&0.00199	&0.00283	
	&0.00353	&0.00419	&0.0049		&0.00525\\
\hline\hline
\end{tabular}
}
}
\vfill\newpage

\centerline{Table~3}
\centerline{Parameters for the Normalized Pad\'e Approximant $[0,4]$
with Constrained Variance}
\vskip 2 true cm
\mbox{
\small{
\begin{tabular}
{| c  | c  | c  | c  | c  | c  | c  | c  | c  | }
\hline
	&		&\LL=1		&\LL=5		&\LL=10		
	&\LL=15		&\LL=20		&\LL=25		&\LL=30\\
\hline\hline
	& \C		&140.62		&353.13		&573.23		
	&590.96		&765.57		&736.57		&719.84\\
$[$m]=$[$1] & \qu		&25.262		&8.965		&6.130		
	&4.807		&3.968		&3.425		&3.049\\
	& \dqu		&0.517		&0.172		&0.112		
	&0.085		&0.067		&0.061		&0.055\\
	&\ss		&-0.00071	&0.00421	&0.00633	
	&0.00564	&0.02711	&0.00749	&0.00832\\
	&\ds		&0.00078	&0.00231	&0.00346	
	&0.00442	&0.00541	&0.00612	&0.00682\\
\hline
	& \C		&368.13$^*$	&305.13		&457.03		
	&552.25		&660.94		&622.24		&585.35\\
$[$m]=$[$2]	& \qu		&24.056		&8.163		&5.403		
	&4.190		&3.395		&2.928		&2.619\\
	& \dqu		&0.475		&0.166		&0.103		
	&0.075		&0.059		&0.054		&0.049\\
	&\ss		&-0.00008	&0.00456	&0.01033	
	&0.01152	&0.01581	&0.00845	&-0.00263\\
	&\ds		&0.00081	&0.00241	&0.00363	
	&0.00469	&0.00568	&0.00659	&0.00740\\
\hline
	& \C		&126.50		&288.11		&352.29		
	&398.49		&550.75		&542.31		&525.85\\
$[$m]=$[$3]	& \qu		&24.260		&7.928		&5.163		
	&3.905		&3.244		&2.816		&2.550\\
	& \dqu		&0.497		&0.163		&0.102		
	&0.073		&0.057		&0.058		&0.052\\
	&\ss		&0.00013	&0.00549	&0.00661	
	&0.00538	&0.01881	&0.01162	&0.00225\\
	&\ds		&0.00082	&0.00242	&0.00363	
	&0.00471	&0.00552	&0.00644	&0.00727\\
\hline
	& \C		&137.99		&205.47		&288.40		
	&336.55		&443.68		&503.88		&502.02\\
$[$m]=$[$5]	& \qu		&24.276		&7.975		&5.328		
	&4.135		&3.429		&2.998		&2.760\\
	& \dqu		&0.509		&0.159		&0.103		
	&0.079		&0.063		&0.056		&0.051\\
	&\ss		&0.00005	&0.00203	&0.00689	
	&0.00251	&0.00088	&0.00006	&-0.00402\\
	&\ds		&0.00081	&0.00240	&0.00349	
	&0.00444	&0.00520	&0.00604	&0.00671\\
\hline
	& \C		&136.33		&172.98		&260.75		
	&335.45		&339.23		&368.04		&516.76\\
$[$m]=$[$7]	& \qu		&23.817		&8.032		&5.601		
	&4.360		&3.547		&3.102		&2.958\\
	& \dqu		&0.490		&0.161		&0.111		
	&0.085		&0.066		&0.059		&0.053\\
	&\ss		&-0.00046	&0.00171	&0.00447	
	&0.00265	&-0.00040	&0.00123	&-0.00896\\
	&\ds		&0.00083	&0.00237	&0.00338	
	&0.00424	&0.00506	&0.00581	&0.00614\\
\hline
	& \C		&153.87		&178.63		&231.94		
	&319.50		&377.44		&325.44		&499.18\\
$[$m]=$[$10]& \qu		&24.630		&8.527		&5.810		
	&4.592		&3.823		&3.286		&3.127\\
	& \dqu		&0.505		&0.169		&0.113		
	&0.091		&0.073		&0.064		&0.058\\
	&\ss		&-0.00077	&0.00189	&0.00722	
	&0.00234	&0.00328	&0.00012	&0.00480\\
	&\ds		&0.00080	&0.00224	&0.00324	
	&0.00400	&0.00474	&0.00556	&0.00581\\
\hline
	& \C		&126.60		&177.63		&248.66		
	&289.85		&346.74		&317.15		&478.07\\
$[$m]=$[$30]& \qu		&27.836		&9.876		&6.888		
	&5.486		&4.536		&3.909		&3.638\\
	& \dqu		&0.578		&0.195		&0.135		
	&0.107		&0.087		&0.075		&0.070\\
	&\ss		&-0.00060	&0.00109	&0.00430	
	&0.00123	&0.00167	&0.00275	&-0.00657\\
	&\ds		&0.00071	&0.00197	&0.00280	
	&0.00349	&0.00412	&0.00481	&0.00516\\
\hline\hline
\end{tabular}
}
}
\vfill\newpage
\centerline{Table~4}
\centerline{Parameters of the Pad\'e $[0,4]$ as a Function of the
Lag for Different Maturities}
\vskip 2 true cm
\mbox{
\small{
\begin{tabular}
{| c  | c  | c  | c  | c  | c  | c  | c | }
\hline
	&\matur{1}	&\matur{2}	&\matur{3}	&\matur{5}	
	&\matur{7}	&\matur{10}	&\matur{30}\\
\hline\hline
\laqu	&-0.663	&-0.666	&-0.660	&-0.623	&-0.604	&-0.594	&-0.582\\
\hline
\dlaqu	&0.004	&0.004	&0.004	&0.004	&0.004	&0.004	&0.004	\\
\hline
\nuqu	&1.480	&1.363	&1.350	&1.411	&1.459	&1.513	&1.676\\
\hline
\dnuqu	&0.004	&0.004	&0.004	&0.004	&0.004	&0.004	&0.004\\
\hline
\laqd	&-1.054	&-1.070	&-1.085	&-1.076	&-1.054	&-1.046	&-1.052\\
\hline
\dlaqd	&0.010	&0.009	&0.009	&0.008	&0.009	&0.008	&0.009\\
\hline
\nuqd	&1.647	&1.684	&1.670	&1.737	&1.803	&1.907	&2.042	\\
\hline
\dnuqd	&0.008	&0.007	&0.007	&0.007	&0.007	&0.007	&0.008\\
\hline\hline
\end{tabular}
}
}

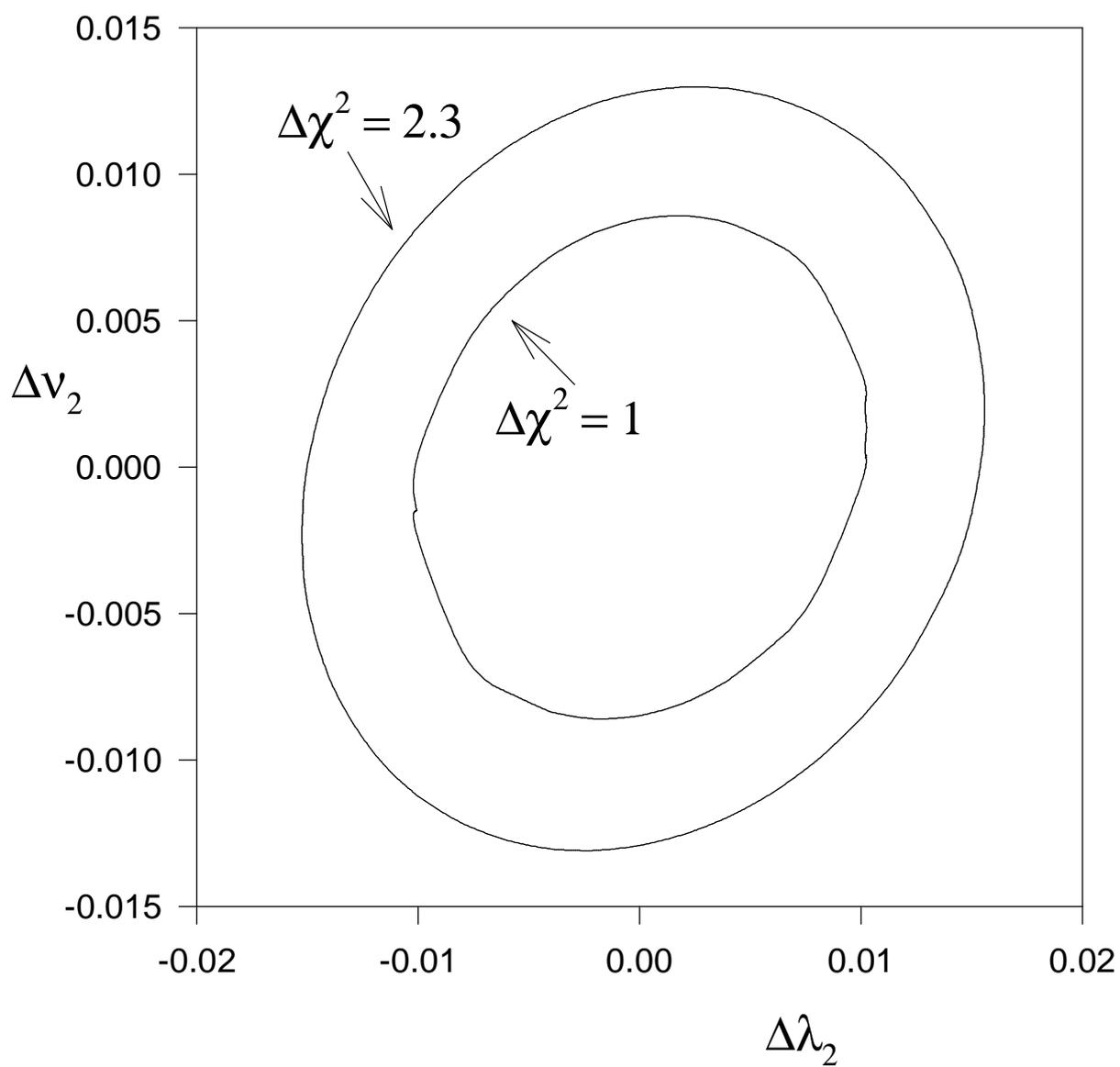
\end{document}